\documentclass[conference]{IEEEtran}
\IEEEoverridecommandlockouts
\usepackage{cite}
\usepackage{amsmath,amssymb,amsfonts}
\usepackage{algorithmic}
\usepackage{graphicx}
\usepackage{textcomp}
\usepackage{xcolor}
\usepackage{subcaption}
\usepackage{hyperref}
\usepackage{tabularx}
\usepackage{float}
\usepackage{soul}
\usepackage{textcomp}
\soulregister\ref7
\soulregister\cite7
\soulregister\textbf7
\renewcommand\hl[1]{#1} 
\usepackage{array}
\def\BibTeX{{\rm B\kern-.05em{\sc i\kern-.025em b}\kern-.08em
    T\kern-.1667em\lower.7ex\hbox{E}\kern-.125emX}}
\begin{document}
\bstctlcite{IEEEexample:BSTcontrol}
\title{Spatiotemporal Analysis of VIIRS Satellite Observations and Network Traffic During the 2025 Manitoba Wildfires\\
\thanks{We acknowledge the support of the Natural Sciences and Engineering Research Council of Canada (NSERC) [funding reference number RGPIN-2022-03364], Research Manitoba, as well as the support provided by Ookla Inc. and Research Manitoba grants.}
}

\author{\IEEEauthorblockN{Xiang Shi and Peng Hu}
\IEEEauthorblockA{
Advanced Network and Embedded Systems Lab (AEL) \\
Dept. of Electrical and Computer Engineering, University of Manitoba, Winnipeg, Canada\\
 shix2@myumanitoba.ca, peng.hu@umanitoba.ca}
}

\maketitle

\begin{abstract}
Climate change has intensified extreme weather and wildfire conditions globally. Canada experienced record-breaking wildfires in 2023 and 2025, burning millions of hectares and severely impacting the Prairie provinces, with Manitoba facing its worst season in 30 years. These events highlight the urgent need to understand and mitigate escalating fire risks. While existing research largely focuses on wildfire management approaches, few studies have explored the relationship between user network traffic and wildfire activity, despite the potential of such correlations to provide valuable spatiotemporal insights into wildfire dynamics. This paper investigates the relationship between wildfire intensity and network performance during the 2025 Manitoba wildfire season, using Visible Infrared Imaging Radiometer Suite (VIIRS) satellite-derived Fire Radiative Power data and large-scale Speedtest measurements. We found statistically significant correlations between wildfire intensity and several network performance metrics in both the province-wide and region-wide case studies, as measured by Spearman's correlation coefficients ($\rho$) and corresponding p-values. Throughput-related metrics showed inverse correlations with wildfire intensity (e.g., download speed: $\rho = -0.214$, $p\_value = 0.004$), whereas latency-related metrics showed positive correlations (e.g., round-trip time latency: $\rho = 0.162$, $p\_value = 0.0308$). The findings suggest satellite fire indicators and network performance metrics together can reveal vulnerabilities during extreme environmental events and support diaster response and recovery efforts.
\end{abstract}


\section{Introduction}
Climate change has significantly intensified extreme weather events and fire-weather conditions in Canada and around the world. In recent years, much of the world has seemed ``on fire.'' The United Nations Secretary-General confirmed that July 2023 was the hottest month on record, with record-breaking global temperatures continuing into 2024. Canada experienced its worst wildfire season on record in 2023, followed by another severe season in 2025, the second worst on record. In 2023 alone, wildfires burned more than 17.2 million hectares, while in 2025 over 8.9 million hectares were affected \cite{b16}, impacting millions of Canadians and numerous industries. Between 2023 and 2025, Canada’s Prairie provinces, particularly Alberta, Saskatchewan, and Manitoba, experienced some of the most intense and destructive wildfires. Manitoba, in particular, endured its worst wildfire season in the past 30 years.

Research efforts have addressed this challenge through a range of technical approaches,  including fire detection platforms, burned-area mapping, and wildfire prediction using satellite-based Earth observations. Among satellite observation solutions, the Visible Infrared Imaging Radiometer Suite (VIIRS) has played an essential role. VIIRS is onboard the Suomi NPP (S-NPP), NOAA-20, and NOAA-21 satellites and provides the multi-band visible and infrared imagery for detecting active fires and providing information on smoke plumes, aerosols, clouds, and land vegetation. However, there has been hardly research efforts in the correlation between user network traffic and wildfire events, even though such correlations may offer important spatiotemporal insights into wildfire occurrence and dynamics. For example, correlation patterns can help analyze digital connectivity gaps and indicate where enhanced, resilient Internet infrastructure may be required in affected regions. In addition, if such a correlation exists, network traffic can serve as a user-oriented, crowd-sourced indicator of the early onset of wildfires, supporting analyses of social impact and community vulnerability as well as informing climate adaptation strategies.


Previous research has mostly explored the impact of wildfires on the natural environments \cite{b13, b14}, while leaving a fundamental question unaddressed: whether wildfire events influence network performance, or conversely, whether network activity is affected by such events. \hl{Beyond establishing whether such relationships exist, it is also essential to identify which network performance metrics are affected.} Motivated by these gaps, the main contributions of the paper are summarized as follows:

\begin{itemize}
\item \hl{We create province-wide and region-wide wildfire intensity datasets using Fire Radiative Power (FRP) observations retrieved from the NASA Fire Information for Resource Management System (FIRMS).}

\item We explore the correlation between network performance metrics and wildfire intensity in Manitoba using rank-based correlations and visual evidence.

\item \hl{Our analysis results show that the wildfire intensity and several network performance metrics exist statistically significant correlations, including upload/download speeds and latency-related metrics, i.e., upload/download latency and latency measured in round-trip time (RTT).}

\end{itemize}

The remainder of the paper is structured as follows. Section II discusses the related work. Section III discusses the geographic areas of the case studies and describes the datasets used. Section IV provides methodology and analytical results. The conclusive remarks are made in Section V.

\section{Related Work}



In recent years, wildfire research has increasingly utilized satellite-based remote sensing tools to acquire wildfire data \cite{b4, b5, b6, b7, b8, b11}. FRP is a quantitative metric used to characterize wildfire intensity \cite{b2}. It is commonly employed as the primary fire intensity metric across domains such as spatial aggregation \cite{b10}, air quality \cite{b13}, air temperature \cite{b14}, infrastructure damage \cite{b12}, and ecological impact \cite{b8, b9}. Urfalı \textit{et al.}\cite{b3} proposed an integrated spatiotemporal framework that combines Spatio-Temporal Density-Based Spatial Clustering of Applications with Noise (ST-DBSCAN) clustering on VIIRS fire detections with FRP and the differenced Normalized Burn Ratio ($\Delta$NBR) to characterize the dynamics and ecological impacts of large-scale wildfires during the extreme 2023 Quebec fire season. They used 80,228 VIIRS fire records, which resulted in 19 distinct clusters across four active fire regions, and reported strong spatial agreement with the National Burned Area Composite (NBAC) dataset. Importantly, by using Gaussian Process Regression (GPR), they found a significant non-linear relationship between FRP and key fire behavior metrics, such that higher mean FRP was associated with longer durations and greater burn severity. These findings support the use of FRP as a meaningful strength variable to represent wildfire intensity. 

There are two popular rank-based correlation methods used to identify the nonlinear relationship between two continuous variables. Rank-based correlation measures association by transforming the data into ranks (i.e., ordered positions) rather than the original values. Spearman's $\rho$ and Kendall's $\tau$ are rank-based and non-parametric measures of correlation between variables $X$ and $Y$. However, their interpretations are different \cite{b15}, as Spearman's ($\rho$) is considered the regular Pearson's correlation coefficient in terms of the proportion of variability accounted for, whereas Kendall's ($\tau$) uses a probability rank difference, i.e., the difference between the probability that the sample data are in the same order versus the probability that the sample data are not in the same order. \hl{Kariuki \textit{et al.}\cite{b17} used the Spearman rank-based correlation method to analyze the relationship between wildfire hotspot densities and weather variables. Their research found that multiple drought propagation factors were strongly and positively correlated with wildfire density in Chyulu Hill and Tsavo West National Parks. Meng \textit{et al.}\cite{b18} utilized Spearman rank-based correlation and Mann-Kendall test to investigate the association between preceding winter snowfall and spring wildfire in Northeastern Asia. Their research concludes that winter snowfall, spring soil surface dryness, and the risk of spring wildfire exhibit a statistically significant correlation. Richter \textit{et al.}\cite{b19} used Pearson, Spearman, and Kendall correlation methods to examine the relationship between transport-layer security (TLS) handshake latency and packet loss in the Starlink network. They found no statistically significant correlation between the two variables. Based on the state of the art, we can see that research on the correlation between wildfires and network traffic has hardly been conducted in the literature.}

\section{Study Area and Dataset Description}
\begin{figure}[htbp]
     \centering
     \begin{subfigure}{0.45\textwidth}
         \centering
         \includegraphics[width=\linewidth]{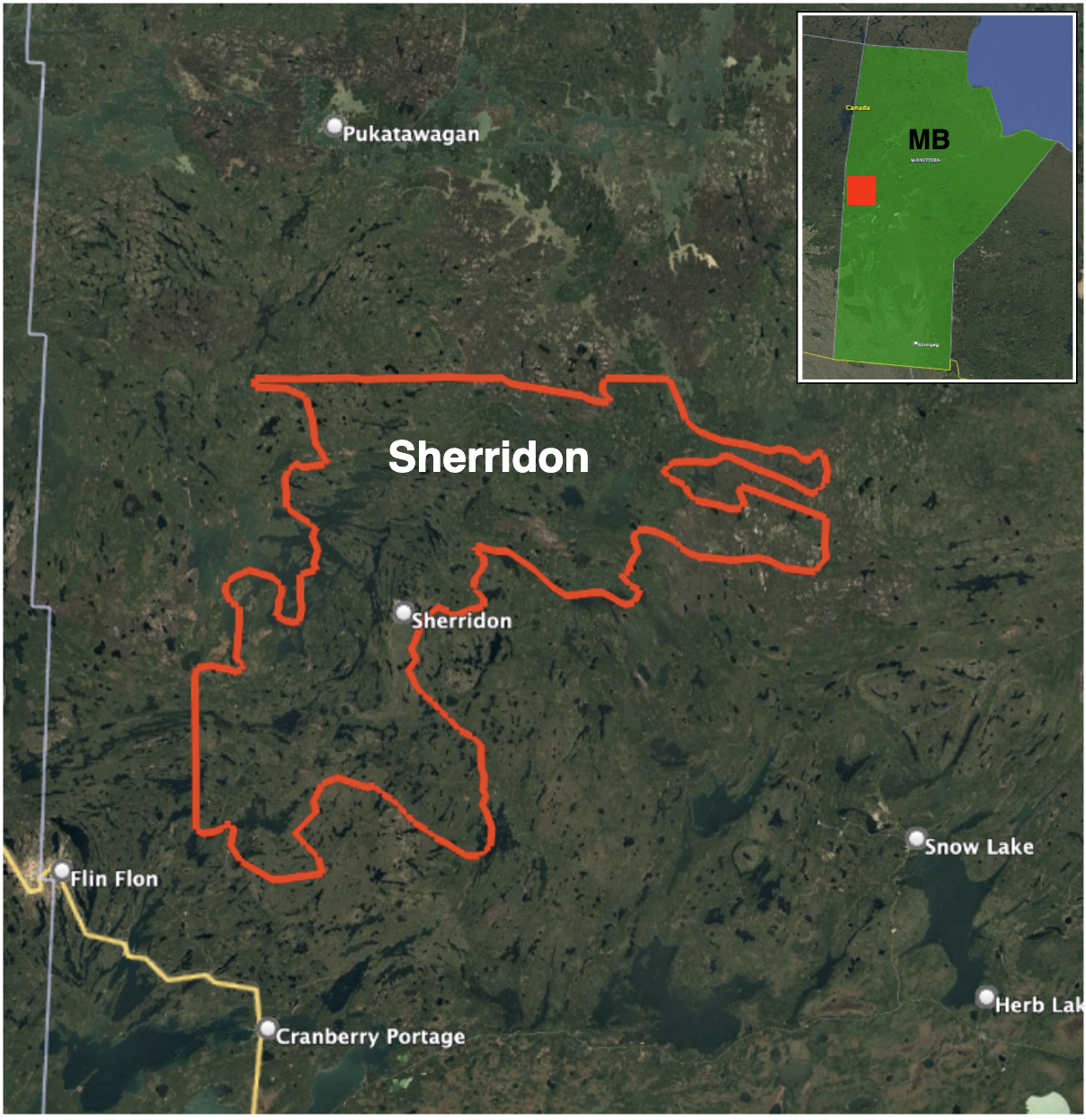}
         \caption{Map of the Sherridon region in Manitoba, Canada.}
         \label{fig:regional_map}
     \end{subfigure}
     \hfill
     \begin{subfigure}{0.45\textwidth}
         \centering
         \includegraphics[width=\linewidth]{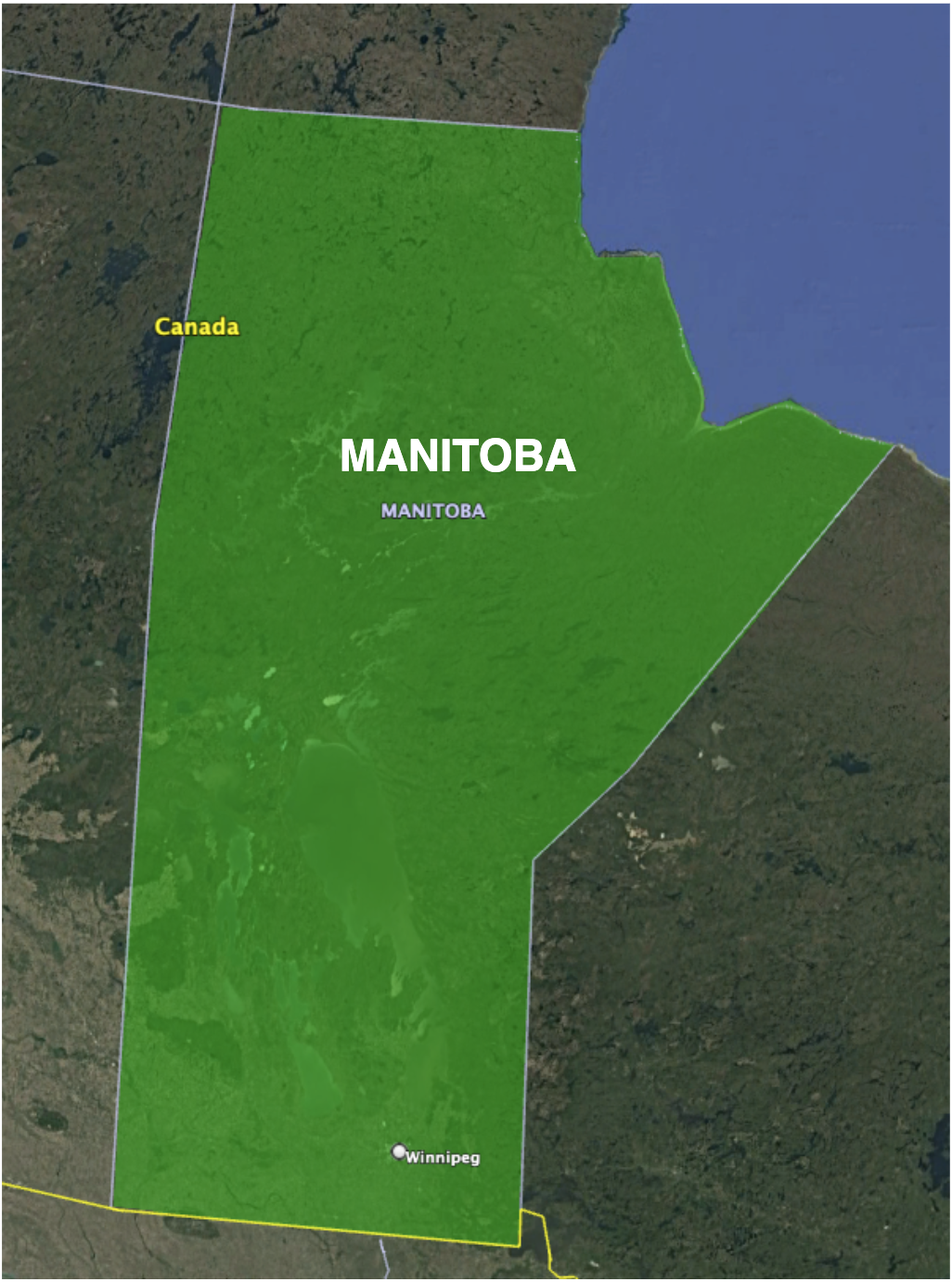}
         \caption{Province-wide map of Manitoba, Canada.} 
         \label{fig:province_map}
     \end{subfigure}
     \caption{Study areas in Manitoba: (a) Sherridon region and (b) Manitoba province. In both maps, the geographical regions are shown with red boundaries.}
     \label{fig:study_are_maps}
\end{figure}

\textbf{Study areas}: As shown in Fig.~\ref{fig:study_are_maps}, our study covers Manitoba at two spatial scales: a regional area of interest around Sherridon (as depicted in Fig.~\ref{fig:regional_map}) and the province-wide boundary (as depicted in Fig.~\ref{fig:province_map}). We select the Sherridon region for our regional case study because it provides sufficient spatiotemporal alignment between the network and wildfire datasets, thereby enhancing the reliability of the correlation analysis.

\textbf{Wildfire dataset:} Wildfire detection datasets were obtained from the NASA FIRMS website (see  \url{https://firms.modaps.eosdis.nasa.gov/}) using VIIRS wildfire detection from S-NPP, NOAA-20, and NOAA-21 satellites. We download the satellite-derived datasets based on the predefined study areas and observation period. While data were retrieved in plain-text tabular format for downstream processing. Datasets contain common features, including acquisition time, latitude/longitude, FRP, brightness, confidence class, and day/night flags.

\textbf{Network measurement dataset:} Our network data is provided by Ookla's Speedtest measurements. Based on the Speedtest methodology (\url{https://www.ookla.com/resources/guides/speedtest-methodology}), the measurement data is categorized into two types: fixed broadband and mobile network connections. Both types of datasets contain abundant network performance metrics, including throughput, latency-related, packet loss rate, jitter, user device information, network provider information, measurement time in coordinated universal time (UTC), and geographic location (i.e., latitude and longitude). For our analysis, we only focus on the fixed broadband dataset. Among the spatiotemporal matching network and wildfire datasets, only the fixed broadband dataset provides a sufficient sample size. More details will be explained in Section IV. 

\section{Spatiotemporal Analysis \& Results}

We discuss spatiotemporal correlation analysis of wildfire intensity and network performance at both the provincial and regional scales in Manitoba over the observation period from May 1, 2025 to Oct 1, 2025. We quantify associations using rank-based correlations methods (e.g., Spearman's $\rho$ and Kendall’s $\tau$) and visual evidence (e.g., scatter and time-series plots).

\subsection{Variables and Metrics for Correlation Analysis}

This paragraph defines the wildfire intensity variable and network performance metrics selected for our case studies. FRP has been utilized as the variable representing wildfire intensity throughout the study. In addition, we use throughput (upload/download speeds measured in Mbps), latency-related (upload/download latency and RTT, all measured in ms), and packet loss as our network performance metrics.

\begin{table}[hbtp]
    \centering
    \caption{Sherridon region rank-based correlation coefficients between network performance metrics and FRP. Here we report Spearman’s $\rho$ and Kendall’s $\tau$ results, along with $p\_value$ and sample size $(N)$.}
    \label{tab:spearman_sherridon}
    \small
    \setlength{\tabcolsep}{4pt}
    \renewcommand{\arraystretch}{0.95}
    \begin{tabularx}{\linewidth}{|>{\raggedright\arraybackslash}X|c|c|c|c|}
        \hline
        \multicolumn{5}{|c|}{\textbf{Spearman’s $\rho$ rank-based correlation}} \\
        \hline
        \textbf{Network Performance Metric} & \textbf{$N$} & \textbf{$\rho$} & \textbf{$p\_value$} & \textbf{significance$^{\mathrm{1}}$} \\
        \hline
        Upload speed$^{\mathrm{2}}$ & 290 & -0.121 & 0.039198 & significant \\
        Download speed$^{\mathrm{3}}$ & 290 & -0.094 & 0.111358 & ns \\
        RTT$^{\mathrm{4}}$ & 263 & 0.097 & 0.115307 & ns \\
        Upload latency$^{\mathrm{5}}$ & 261 & 0.195 & 0.001586 & significant \\
        Download latency$^{\mathrm{6}}$ & 261 & 0.116 & 0.061309 & ns \\
        PL$^{\mathrm{7}}$ & 190 & 0.056 & 0.443682 & ns\\
        \hline
    \end{tabularx}
    
    \vspace{2em}
    
    \label{tab:kendall_sherridon}
    \small
    \setlength{\tabcolsep}{4pt}
    \renewcommand{\arraystretch}{0.95}
    \begin{tabularx}{\linewidth}{|>{\raggedright\arraybackslash}X|c|c|c|c|}
        \hline
        \multicolumn{5}{|c|}{\textbf{Kendall’s $\tau$ rank-based correlation}} \\
        \hline
        \textbf{Network Performance Metric} & \textbf{$N$} & \textbf{$\tau$} & \textbf{$p\_value$} & \textbf{significance$^{\mathrm{1}}$} \\
        \hline
        Upload speed$^{\mathrm{2}}$ & 290 & -0.079 & 0.048765 & significant \\
        Download speed$^{\mathrm{3}}$ & 290 & -0.060 & 0.129858 & ns \\
        RTT$^{\mathrm{4}}$ & 263 & 0.069 & 0.098069 & ns \\
        Upload latency$^{\mathrm{5}}$ & 261 & 0.131 & 0.001777 & significant \\
        Download latency$^{\mathrm{6}}$ & 261 & 0.077 & 0.066268 & ns \\
        PL$^{\mathrm{7}}$ & 190 & 0.039 & 0.493304 & ns \\
        \hline
        \multicolumn{5}{p{\dimexpr\linewidth-2\tabcolsep\relax}}
        {\footnotesize{$^{\mathrm{1}}$Significance codes are defined by us whether the correlation is statistically significant, corresponding codes: not significant (ns) $p\_value \ge 0.05$, and significant is $p\_value < 0.05$. 
        $^{\mathrm{2}}$upload speed in Mbps.
        $^{\mathrm{3}}$download speed in Mbps.
        $^{\mathrm{4}}$RTT in ms.
        $^{\mathrm{5}}$upload latency in ms.
        $^{\mathrm{6}}$download latency in ms.
        $^{\mathrm{7}}$packet loss percent.}}\\ 
    \end{tabularx}
\end{table}

\vspace{-5pt}
\subsection{Sherridon Region Case Study}

First, we match the network dataset to FRP records based on time and distance variables. In the Sherridon region, we obtained 290 matched samples by linking each network data to the nearest FRP record within 24 hours and 50 km.

To quantify the association between FRP and network performance metrics in the regional case study, we compute Spearman's $\rho$ and Kendall's $\tau$ rank-based correlation coefficients and their p-values over the observation period. The results are summarized in Table~\ref{tab:spearman_sherridon}. Based on Spearman's and Kendall's correlation results, we identify that FRP is statistically significantly correlated with the network's upload speed and upload latency. For the upload speed, both rank-based correlation coefficients are negative (i.e., Spearman's $\rho = -0.121$, $p\_value = 0.039198$; Kendall's $\tau = -0.079$, $p\_value = 0.048765$), indicating an inverse monotonic relationship with FRP, meaning that as FRP increases, upload speed decreases. In contrast, the upload latency is positively correlated with FRP (i.e., Spearman's $\rho = 0.195$, $p\_value = 0.001586$; Kendall's $\tau = 0.131$, $p\_value = 0.001777$), indicating that higher wildfire intensity tends to coincide with increased latency of upload.

\begin{figure}[t!]
     \centering
     \begin{subfigure}{\columnwidth}
         \centering
         \includegraphics[width=0.95\linewidth, height=5cm]{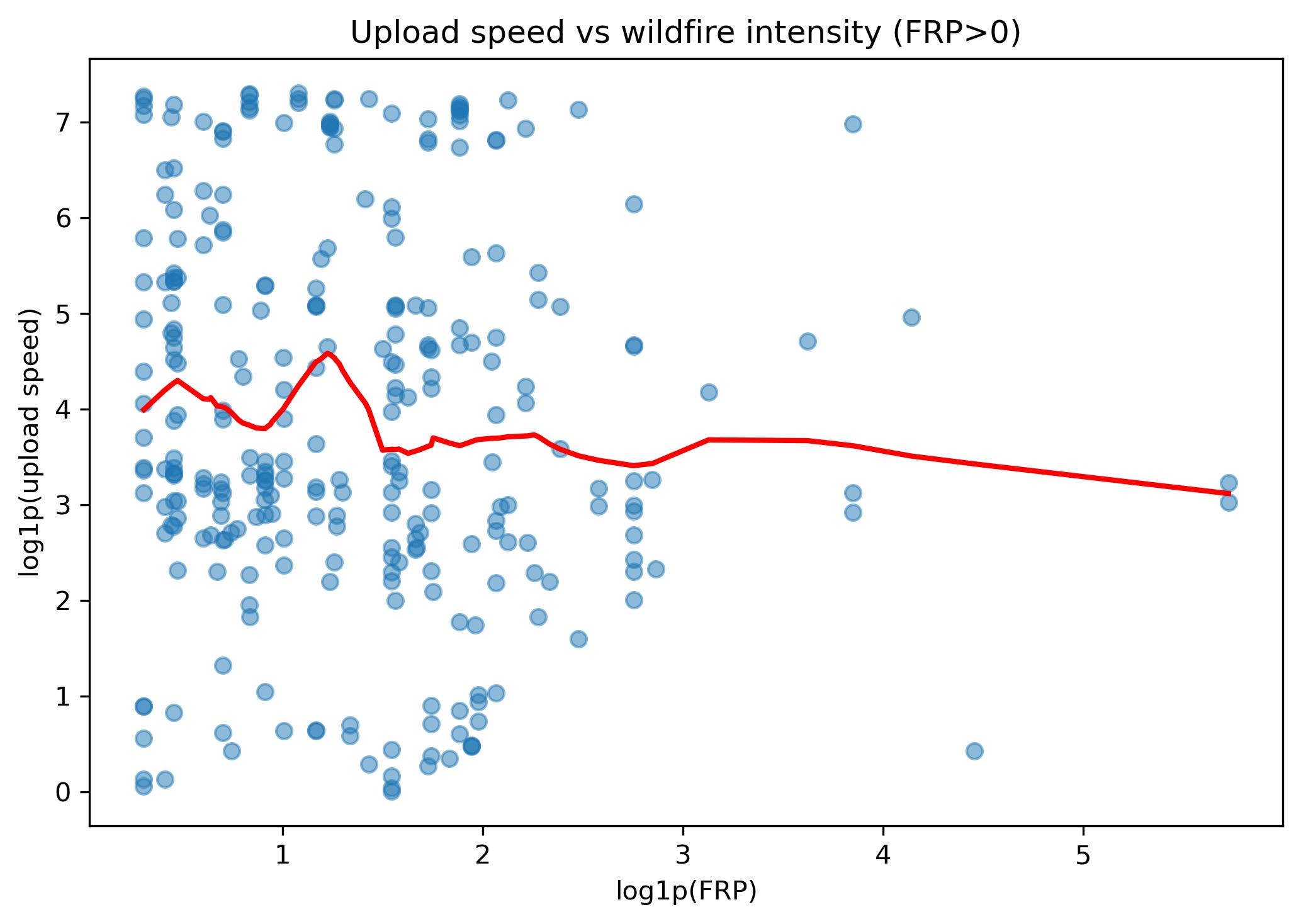}
         \caption{Scatter LOESS plot for upload speed and FRP.} 
         \label{fig:scatter_sub_upload}
     \end{subfigure}
     \hfill
     \begin{subfigure}{\columnwidth}
         \centering
         \includegraphics[width=0.95\linewidth, height=5cm]{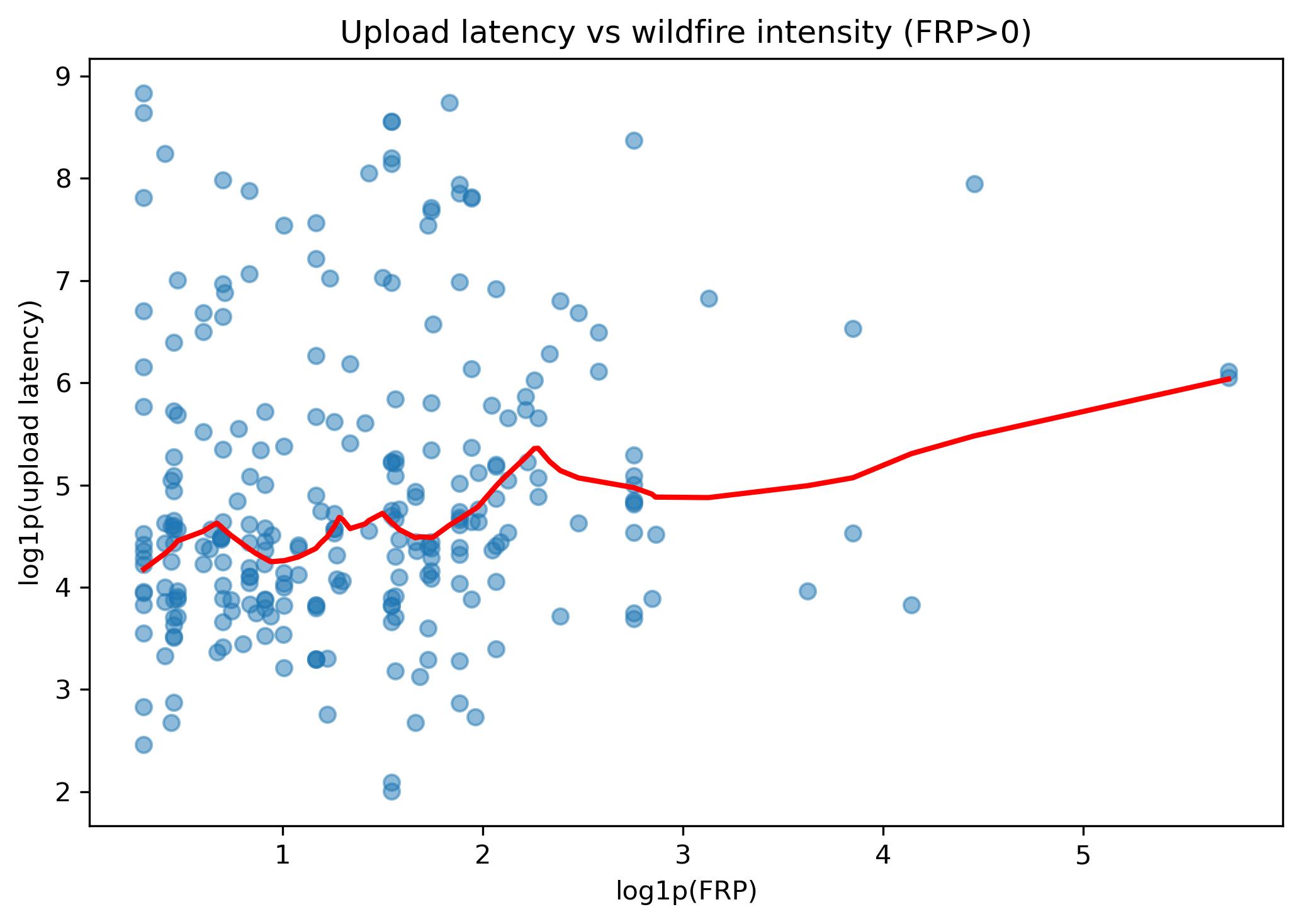}
         \caption{Scatter LOESS plot for upload latency and FRP.}
         \label{fig:scatter_sub_upload_latency}
     \end{subfigure}
    \caption{Scatter LOESS plots for upload speed/latency versus FRP in the Sherridon region. Both horizontal and vertical axes are log1p-transformed.}
     \label{fig:scatter_plots_for_upload_speed_latency}
\end{figure}

\begin{figure}[htp]
     \centering
     \begin{subfigure}{\columnwidth}
         \centering
        \includegraphics[width=0.98\linewidth, height=4cm]{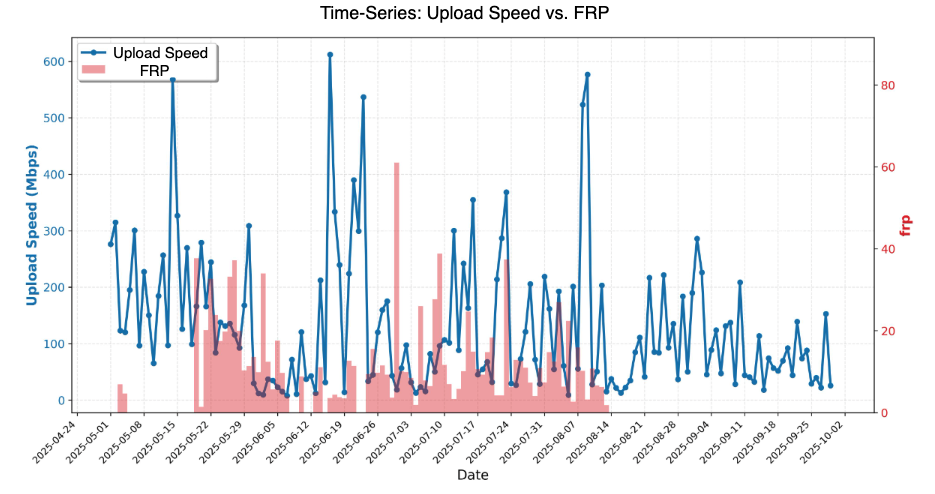}
         \caption{Time-series plot of upload speed and FRP.} 
         \label{fig:ts_upload_speed}
     \end{subfigure}
     \hfill
     \begin{subfigure}{\columnwidth}
         \centering
         \includegraphics[width=0.98\linewidth, height=4cm]{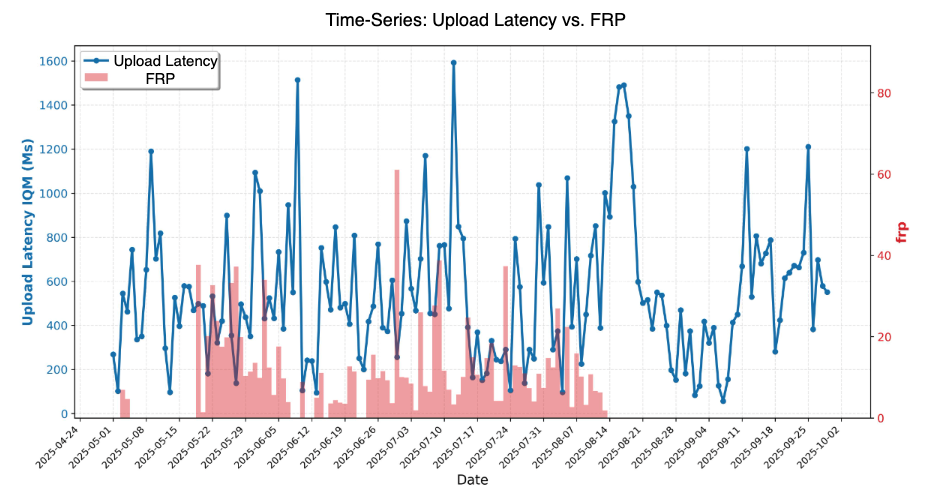}
         \caption{Time-series plot of upload latency and FRP.}
         \label{fig:ts_upload_latency}
     \end{subfigure}
     \caption{Time-series plots of network upload speed and upload latency with FRP in the Sherridon region. The blue line represents upload speed or latency, while the red bars indicate average FRP values.}
     \label{fig:ts_upload_speed_and_latency}
\end{figure}

To visually examine the relationship between FRP and upload speed/latency, we further plot log1p-transformed scatter plots with LOESS-smoothed trend lines in Fig.~\ref{fig:scatter_plots_for_upload_speed_latency}. To mitigate the skewness of the data distribution, we leveraged the $log1p$ transformation~\ref{equ:log1p-transformation} for all variables.

\begin{equation}
    \label{equ:log1p-transformation}
    f(x) = \ln(1 + x)
\end{equation}

In these plots, the horizontal axes represent transformed FRP, while the vertical axes represent the transformed upload speed (Fig.~\ref{fig:scatter_sub_upload}) and upload latency (Fig.~\ref{fig:scatter_sub_upload_latency}), respectively. Each point represents one matched sample over the observation period. The red curve is a LOESS-smoothed trend line summarizing the overall relationship between FRP and the corresponding network performance metrics. 


Aligning with the quantitative results, Fig.~\ref{fig:scatter_sub_upload} shows that upload speed decreases slightly as FRP increases, indicating a negative association between upload speed and FRP. In addition, Fig.~\ref{fig:scatter_sub_upload_latency} shows an increasing trend in upload latency with higher FRP values, suggesting a positive association between them. Furthermore, time-series plots between upload speed/latency and FRP are highly consistent with our findings, as shown in Fig.~\ref{fig:ts_upload_speed_and_latency}, providing another visual perspective on the network performance under both fire and non-fire conditions. Overall, the visualization results are consistent with our quantitative findings. 

\subsection{Province-Wide Case Study}
For the province-wide case study, since there is a sufficient amount of measurement data, we apply a smaller matching tolerance (i.e., one hour and five kilometers) to obtain more precise results. Totally 179 matched samples have been obtained after data matching. 

\begin{table}[htbp]
    \centering
    \caption{Province-wide rank-based correlation coefficients between network performance metrics and FRP. Here we report Spearman’s $\rho$ and Kendall’s $\tau$, along with $p\_value$ and sample size $(N)$.}
    \label{tab:mb_corr}
    \small
    \setlength{\tabcolsep}{4pt}
    \renewcommand{\arraystretch}{0.95}
    \begin{tabularx}{\linewidth}{|>{\raggedright\arraybackslash}X|c|c|c|c|}
        \hline
        \multicolumn{5}{|c|}{\textbf{Spearman’s $\rho$ rank-based correlation}} \\
        \hline
        \textbf{Network Performance Metric} & \textbf{$N$} & \textbf{$\rho$} & \textbf{$p\_value$} & \textbf{significance$^{\mathrm{1}}$} \\
        \hline
        Upload speed$^{\mathrm{2}}$ & 179 & -0.087 & 0.245002 & ns \\
        Download speed$^{\mathrm{3}}$ & 179 & -0.214 & 0.004008 & significant \\
        RTT$^{\mathrm{4}}$ & 178 & 0.162 & 0.030843 & significant\\
        Upload latency$^{\mathrm{5}}$ & 176 & 0.119 & 0.116914 & ns \\
        Download latency$^{\mathrm{6}}$ & 176 & 0.230 & 0.002140 & significant\\
        PL$^{\mathrm{7}}$ & 133 & 0.160 & 0.065659 & ns \\
        \hline
    \end{tabularx}

    \vspace{2em}

    \small
    \setlength{\tabcolsep}{4pt}
    \renewcommand{\arraystretch}{0.95}
    \begin{tabularx}{\linewidth}{|>{\raggedright\arraybackslash}X|c|c|c|c|}
        \hline
        \multicolumn{5}{|c|}{\textbf{Kendall’s $\tau$ rank-based correlation}} \\
        \hline
        \textbf{Network Performance Metric} & \textbf{$N$} & \textbf{$\tau$} & \textbf{$p\_value$} & \textbf{significance$^{\mathrm{1}}$} \\
        \hline
        Upload speed$^{\mathrm{2}}$ & 179 & -0.055 & 0.280348 & ns \\
        Download speed$^{\mathrm{3}}$ & 179 & -0.139 & 0.006016 & significant \\
        RTT$^{\mathrm{4}}$ & 178 & 0.106 & 0.037724 & significant \\
        Upload latency$^{\mathrm{5}}$ & 176 & 0.076 & 0.137935 & ns\\
        Download latency$^{\mathrm{6}}$ & 176 & 0.159 & 0.001860 & significant\\
        PL$^{\mathrm{7}}$ & 133 & 0.126 & 0.067313 & ns\\
        \hline
        \multicolumn{5}{p{\dimexpr\linewidth-2\tabcolsep\relax}}
        {\footnotesize{$^{\mathrm{1}}$Significance codes are defined by us whether the correlation is statistically significant, corresponding codes: not significant (ns) $p\_value \ge 0.05$, and significant is $p\_value < 0.05$. 
        $^{\mathrm{2}}$upload speed in Mbps.
        $^{\mathrm{3}}$download speed in Mbps.
        $^{\mathrm{4}}$RTT in ms.
        $^{\mathrm{5}}$upload latency in ms.
        $^{\mathrm{6}}$download latency in ms.
        $^{\mathrm{7}}$packet loss percent.}}\\ 
    \end{tabularx}
\end{table}

\begin{figure}[h]
     \centering
     \begin{subfigure}{\columnwidth}
         \centering
         \includegraphics[width=0.95\linewidth, height=5cm]{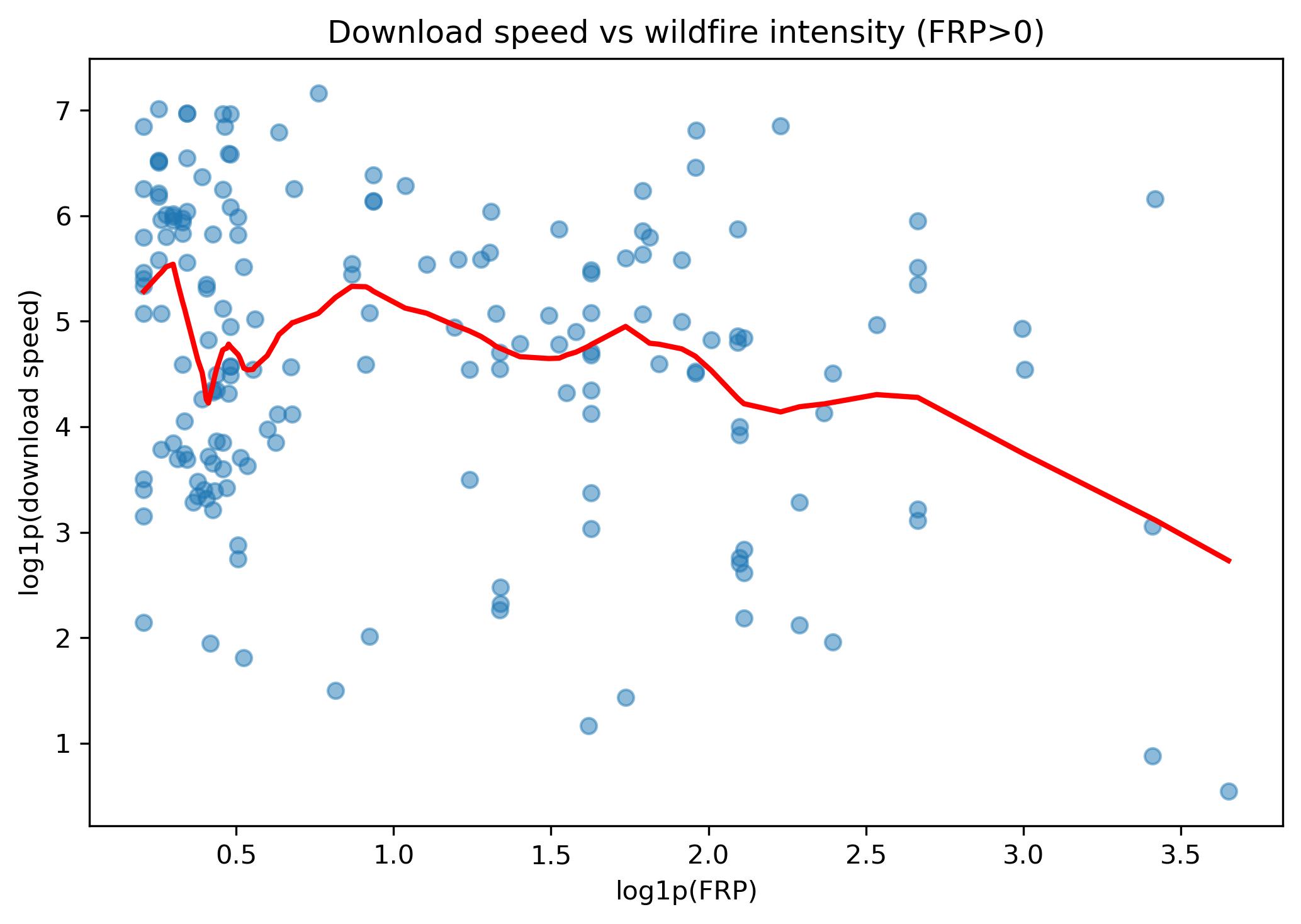}
         \caption{Scatter LOESS plot for download speed and FRP.} 
         \label{fig:scatter_sub_download}
     \end{subfigure}
     \hfill
     \begin{subfigure}{\columnwidth}
         \centering
         \includegraphics[width=0.95\linewidth, height=5cm]{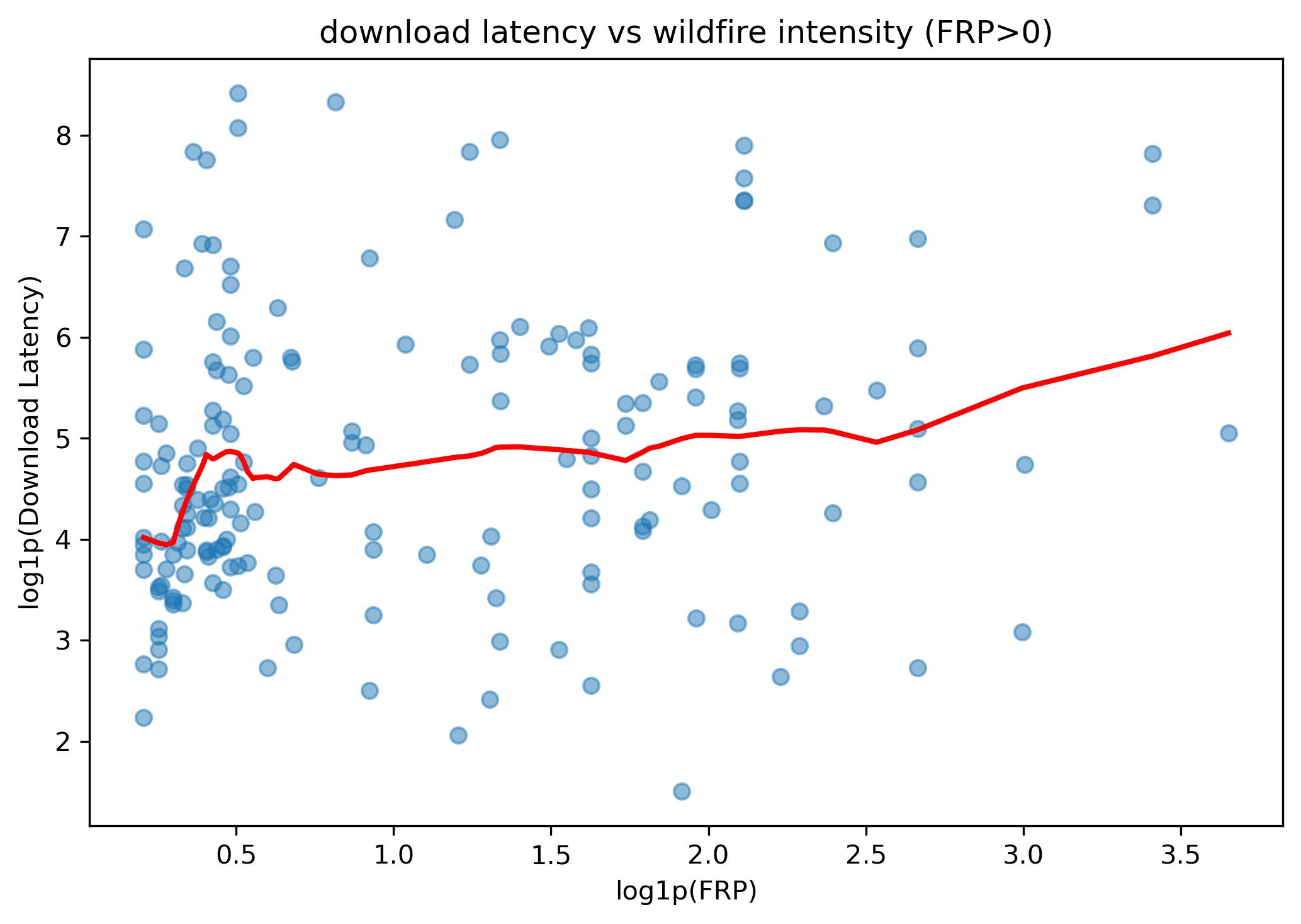}
         \caption{Scatter LOESS plot for download latency and FRP.}
         \label{fig:scatter_sub_download_latency}
     \end{subfigure}
     \hfill
     \begin{subfigure}{\columnwidth}
         \centering
         \includegraphics[width=0.95\linewidth, height=5cm]{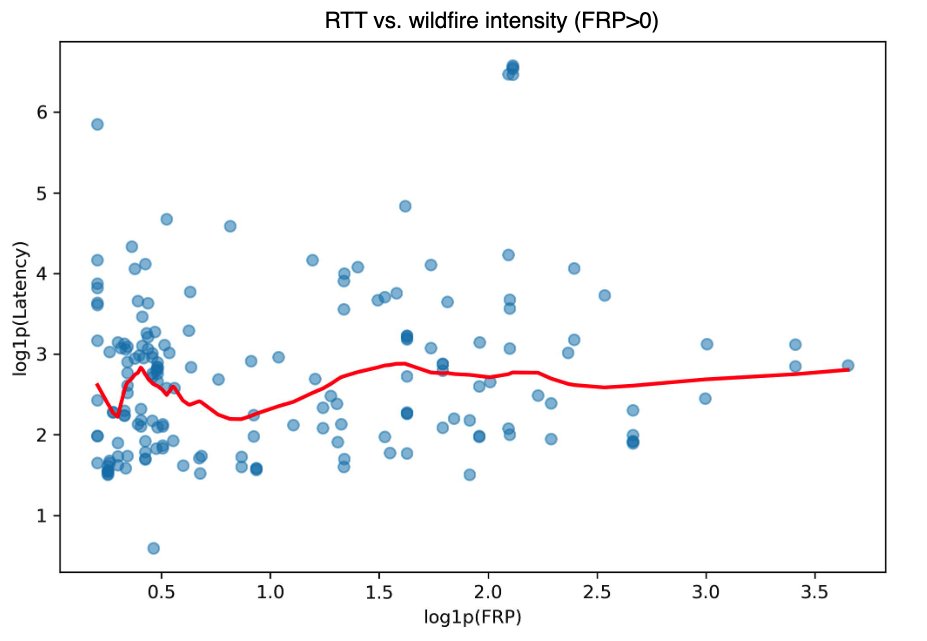}
         \caption{Scatter LOESS plot for RTT and FRP.}
         \label{fig:scatter_sub_latency}
     \end{subfigure}
    \caption{Scatter LOESS plots for download speed, download latency, and RTT vs. FRP in the Province-wide. Both axes in all plots are log1p-transformed.}
     \label{fig:scatter_MB}
\end{figure}

\begin{figure}[tbhp]
     \centering
     \begin{subfigure}{\columnwidth}
         \centering
         \includegraphics[width=0.98\linewidth, height=4cm]{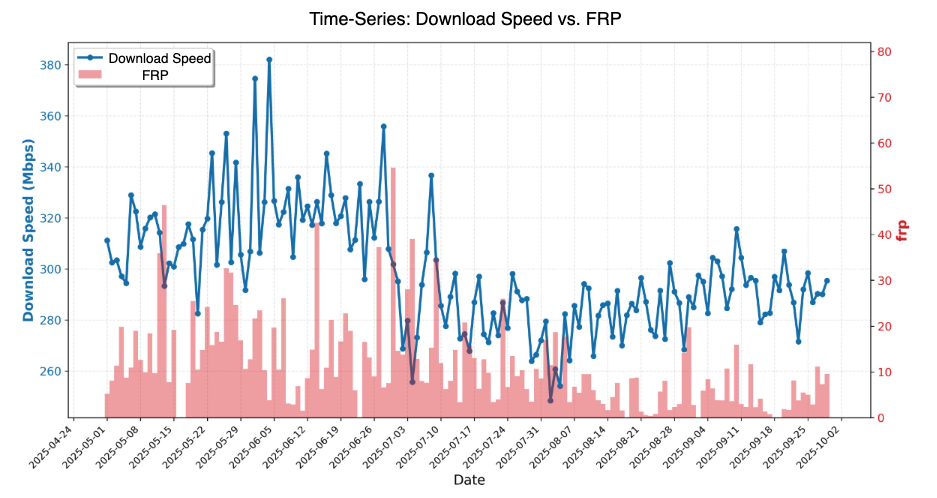}
         \caption{Time-series plot of download speeds and FRP.} 
         \label{fig:ts_mb_download_speed}
     \end{subfigure}
     \hfill
     \begin{subfigure}{\columnwidth}
         \centering
         \includegraphics[width=0.98\linewidth, height=4cm]{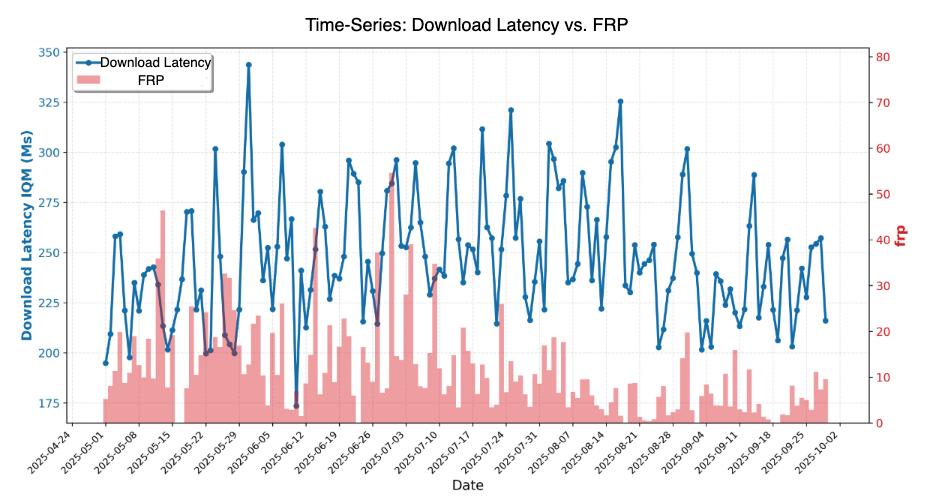}
         \caption{Time-series plot of download latency and FRP.}
         \label{fig:ts_mb_download_latency}
     \end{subfigure}
     \hfill
     \begin{subfigure}{\columnwidth}
         \centering
         \includegraphics[width=0.98\linewidth, height=4cm]{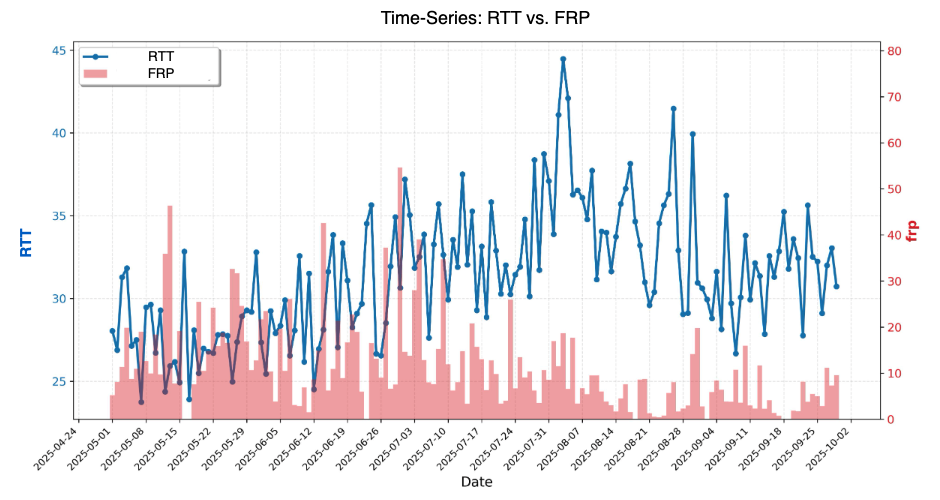}
         \caption{Time-series plot of RTT and FRP.}
         \label{fig:ts_mb_latency}
     \end{subfigure}
     \caption{Time-series plots of network performance metrics (e.g., download speed (Mbps), download latency, and RTT.) with FRP values across the entire Manitoba province. The blue line represents network performance, while the red bars indicate average FRP values.}
     \label{fig:ts_mb}
\end{figure}

Table~\ref{tab:mb_corr} reports the Spearman's $\rho$ and Kendall's $\tau$ rank correlation coefficients results between FRP and network performance metrics. From these results, the FRP exhibits statistically significant correlations with three network performance metrics, including download speed, download latency, and RTT. Download speed shows negative correlation coefficients with FRP under both Spearman's $\rho$ and Kendall's $\tau$. This indicates an inverse monotonic relationship where higher wildfire intensity tends to coincide with lower download speed. In contrast, both download latency and RTT exhibit positive association with FRP, i.e., higher FRP values are generally accompanied by increased delay. In addition, we provide log1p-transformed scatter plots with the LOESS-smoothed trend lines to visualize the relationship between the three network performance metrics and FRP in (Fig.~\ref{fig:scatter_MB}). These three plots are highly consistent with our quantitative results. Fig.~\ref{fig:scatter_sub_download} shows a strong decreasing trend as the FRP values increased. Fig.~\ref{fig:scatter_sub_download_latency} and Fig.~\ref{fig:scatter_sub_latency} both exhibit that as the FRP value increases, download latency and RTT also increase in stepwise manners. The visualization results align with our quantitative results. Time-series visualization plots are also provided in Fig.~\ref{fig:ts_mb} for another analysis perspective.

From the results, we can see the observed effect sizes in upload and download latencies are particularly critical for real-time interactive applications. During periods of severe wildfires, the latency performance reached severely degraded levels, with upload latency peaking at around 1500~ms, and download latency increasing to 350~ms. Such degradation of network quality shifts network communication to unreliable or non-functional. Specifically, the ITU-T G.114 standard \cite{ITU_G114} defines a one-way latency of 400 ms as the upper limit for acceptable interactive communication services. In this context, both upload and download latencies exceed or approach the acceptable latency upper limit, which can have fatal consequences in emergency rescue operations or disaster warnings. 

\section{Conclusion}
In this paper, we discusses whether wildfire activity is associated with network performance metrics by integrating NASA FIRMS VIIRS fire detection records with the Speedtest network measurement datasets through spatiotemporal aggregation. Across both region- and province-wide case studies, multi-dimensional evaluations (e.g., rank-based correlation analysis, LOESS-based, and time-series visualizations) consistently indicate that wildfire intensity is linked to five network performance metrics degradation: throughput (upload/download speeds) tends to decrease, while latency-related (upload latency, download latency, and RTT) tend to increase as wildfire intensity rises. These findings suggest that severe wildfire conditions can coincide with degraded network quality and motivate future work, such as moving beyond correlation by applying causal inference designs to test whether wildfire activity causes network performance degradation. The results could contribute to informed decision-making and the development of resilient communication infrastructure to support diaster response and recovery efforts during wildfires.


\bibliographystyle{IEEEtran}
\bibliography{references}

\end{document}